\documentclass[12pt,a4paper]{article}

\usepackage{amsmath}
\usepackage{amssymb}
\usepackage{mathrsfs}
\usepackage{graphicx}
\usepackage{dsfont}
\usepackage[pdftex,colorlinks=true,linkcolor=black,citecolor=black,urlcolor=black]{hyperref}
\usepackage{enumerate}

\numberwithin{equation}{section}

\def\eqn#1{ \begin{eqnarray} #1 \end{eqnarray} }
\def\eq#1 { \begin{equation} #1 \end{equation} }
\def\eqn#1{ \begin{eqnarray} #1 \end{eqnarray} }

\def\a{\alpha}
\def\b{\beta}
\def\e{\epsilon}

\def\d{\partial}
\def\cL{\mathcal{L}}

\def\cD{\mathcal{D}}
\def\cO{\mathcal{O}}

\def\cS{\mathcal{S}}

\def\sl2r{SL(2,\mathbb{R})}

\def\ce{\varepsilon}

\def\C#1{\left\langle #1 \right\rangle}
\newcommand{\lsim}{\mathrel{\hbox{\rlap{\lower.55ex \hbox{$\sim$}} \kern-.3em \raise.4ex \hbox{$<$}}}}
\newcommand{\gsim}{\mathrel{\hbox{\rlap{\lower.55ex \hbox{$\sim$}} \kern-.3em \raise.4ex \hbox{$>$}}}}

\newcommand{\D}{{\nabla}}

 \newcommand{\be}{\begin{equation}}
\newcommand{\ee}{\end{equation}}

\newcommand\Dlr{\raisebox{0.1em}{$\stackrel{\scriptstyle\leftrightarrow}D$}}

\usepackage{bm}

\begin{document}

\title{\begin{flushright}\vspace{-1in}
       \mbox{\normalsize  EFI-14-19}
       \end{flushright}
       \vskip 20pt
The Euler current and relativistic parity odd transport}

\date{\today}

\author{
Siavash Golkar
  \thanks{\href{mailto:golkar@uchicago.edu}
    {golkar@uchicago.edu}},~
   Matthew M. Roberts
   \thanks{\href{mailto:matthewroberts@uchicago.edu}
     {matthewroberts@uchicago.edu}},~
   Dam T. Son
   \thanks{\href{mailto:dtson@uchicago.edu}
     {dtson@uchicago.edu}} \\ \\
   {\it \it Kadanoff Center for Theoretical Physics and Enrico Fermi Institute,}\\
   {\it   University of Chicago, 5640 South Ellis Ave., Chicago, IL 60637 USA}
} 

\maketitle

\begin{abstract}
For a spacetime of odd dimensions endowed with a unit vector field,
we introduce a new topological current that is identically conserved and whose charge is equal to the Euler character of the even dimensional spacelike foliations. The existence of this current allows us to introduce new Chern-Simons-type
terms in the effective field theories describing relativistic quantum
Hall states and (2+1) dimensional superfluids.  Using effective field
theory, we calculate various correlation functions and identify
transport coefficients.  In the quantum Hall case, this current provides the natural relativistic generalization of the Wen-Zee term, required to characterize the shift and Hall viscosity in quantum Hall systems. For the superfluid case this term is required to have nonzero Hall viscosity and to describe superfluids with non s-wave pairing.
\end{abstract}
\newpage
\tableofcontents

\section{Introduction}
\label{sec:intro}

Low energy effective field theories (EFTs) are a natural organizing
principle in physics. They have had great success in describing a wide
range of phenomena, from the classic particle theory example of the
chiral Lagrangian~\cite{Weinberg:1978kz} to an EFT for
quantum Hall systems~\cite{Wen:1992uk,Wen:1992ej}. The primary
ingredients in constructing an effective theory are energy scales,
degrees of freedom, and symmetries (spacetime, global, and gauge). In
particular some of these local symmetries are only preserved up to a
total derivative. Terms in the effective Lagrangian which are only
gauge invariant up to integration by parts, often called Wess-Zumino
terms, are a useful tool in classifying these EFTs and often inform us
of the topological nature of the matter at hand.

Consider quantum Hall (QH) systems as an example. In the
nonrelativistic case the topological nature is dictated by the
Chern-Simons couplings in the effective field theory: a
electromagnetic (standard) Chern-Simons term
$\epsilon^{\mu\nu\rho}A_\mu\partial_\nu A_\rho $ as well as a mixed
Chern-Simons term $\epsilon^{\mu\nu\rho}A_\mu\partial_\nu \omega_\rho$
($\omega$ being the spin connection of the spatial manifold), the
latter often called the Wen-Zee coupling.  These two terms dictate the
Hall conductivity, the Hall viscosity, and the shift~\cite{Wen:1992ej}.  
For systems with Galilean symmetry, an
extension of of these terms to a full EFT of quantum Hall based on
nonrelativistic diffeomorphism invariance~\cite{Son:2005rv} was worked
out in Refs.~\cite{Hoyos:2011ez,Son:2013rqa}. (The third Chern-Simons term
involving only the spin connection, $\omega \wedge d\omega$, has been
recently argued to be connected to the chirality of the edge
modes~\cite{Gromov:2014gta}).

Recently both integer \cite{Novoselov:2005kj,Zhang:2005zz} and
fractional \cite{FQHE-graphene1,QHE-graphene2} quantum Hall states
have been observed in graphene.  Instead of Galilean invariance,
graphene, with its four massless Dirac modes, exhibits an approximate
relativistic invariance\footnote{While the massless Dirac cones have a
  Lorentz symmetry the sound speed is not the speed of light. The
  nearly instantaneous Coulomb interactions breaks the symmetry,
  leading to a relatively slow running of the fermion
  velocity~\cite{Gonzalez:1993uz,Son:2007ja}.}, hence to undertand the
EFT for quantum Hall states in graphene one needs to understand how to
write down topologically invariant terms.  While the standard
Chern-Simons term is Lorentz invariant, the mixed Wen-Zee term
presents a problem since the (2+1) spin connection $\omega_\mu$ is a nonabelian connection, unlike the 1+1 dimensional version.  Recently, we have found a
relativistic version of the Wen-Zee term~\cite{Golkar:2014}); in this
work we expand on the treatment of Ref.~\cite{Golkar:2014} and extend
the formalism to superfluids.

Relativistic superfluids at zero temperature can also be described via
an EFT~\cite{Son:2002} which, in its standard form, contains one
massless degree of freedom: the Goldstone mode from spontaneous
symmetry breaking.  This formalism, as we will see, fail to deliver a
full treatment of parity broken effects in superfluids, in particular
it misses the possibility of a Hall viscosity.  No Wess-Zumino term was
found in (2+1) dimensional relativistic superfluid EFT~\cite{Nicolis:2014}.
On the other hand, the Hall viscosity can be incorporated into an 
effective field theory of parity-odd nonrelativistic 
superfluids~\cite{Hoyos:2013eha}.  This construction also relies on an
abelian spin connection which does not exist in the relativistic case.

In this paper we attack the problems of relativistic quantum Hall and
relativisic superfluids with the introduction of a new Lorentz
covariant conserved current. This current, defined for odd-dimensional
fluids, depends only on the velocity of the fluid and the spacetime
metric. The corresponding conserved charge is the Euler character of
even-dimensional spacelike foliations. Due to this nature we call it
the ``Euler current'' in analogy with the Euler density. Coupling this
current to the electromagnetic connection in the quantum Hall EFT
provides a natural Lorentz invariant generalization of the Wen-Zee
coupling and gives a nonzero Hall viscosity and shift. In the case of
the superfluid the appropriate coupling is most natural in the dual
description where the Goldstone mode is described by a gauge
field. With this coupling we find the requisite Wess-Zumino term for
nonzero Hall viscosity in the superfluid.

The outline of this paper is as follows. In section \ref{sec:current} we define this new current and discuss its various properties. In section \ref{sec:QH} we discuss coupling the Euler current to the relativistic quantum Hall EFT, summarizing and expanding on the results of \cite{Golkar:2014}. In section \ref{sec:SF} we discuss the dual description of the superfluid EFT and its coupling to the Euler current. We conclude in section \ref{sec:conclusions}. We include for completeness the full list of superfluid correlation functions in appendix 
 \ref{App:SF}.
 
\section{The Euler current}
\label{sec:current}

In this section we define the new topological current and look at its various properties and generalizations to other dimensions.

\subsection{Definition and conservation}
\label{subsec:def_cons}

Let us consider a three dimensional manifold $\mathcal{M}$ supplied with a metric $g_{\mu\nu}$ and a vector field $u^\mu$ of unit norm $u_\mu u^\mu = \sigma$. Here $\sigma=\pm 1$ corresponding to positive or negative signature. We consider the following current:
\begin{equation}
\label{eq:top_current_defined}
J^\mu =\frac{1}{8\pi} \varepsilon ^{\mu\nu\rho} \varepsilon^{\alpha\beta\gamma} u_\alpha
		\left( \nabla_\nu u_\beta \nabla_\rho u_\gamma 
		+ \frac\sigma2 R_{\nu\rho\beta\gamma}\right),
\end{equation}
where we define the totally antisymmetric tensor with $|\text{det }g|^{\frac12}\varepsilon^{txy}= +1$. To show the conversation of this current, we note that since $u^\mu$ has constant norm, $u^\mu\nabla_\nu u_\mu=0$ for any direction $\nu$. Hence $\nabla_\nu u_\mu$ is at each point constrained in the two dimensional surface perpendicular to $u$. Keeping this in mind, we define the parallel and perpendicular projectors 
${P_\parallel}^\mu_\nu = \sigma u^\mu u_\nu$ and ${P_\perp}^\mu_\nu=\delta^\mu_\nu-\sigma u^\mu u_\nu$ and  we calculate:
\begin{align}
\label{eq:current_conservation}
8\pi \nabla_\mu J^\mu = &\varepsilon ^{\mu\nu\rho} \varepsilon^{\alpha\beta\gamma} 
	\nabla_\mu u_\alpha \nabla_\nu u_\beta \nabla_\rho u_\gamma 
	+ 2 \varepsilon ^{\mu\nu\rho} \varepsilon^{\alpha\beta\gamma} 
	u_\alpha \nabla_\mu \nabla_\nu u_\beta \nabla_\rho u_\gamma 
	+ \frac\sigma2 \varepsilon ^{\mu\nu\rho} \varepsilon^{\alpha\beta\gamma}
	 \nabla_\mu u_\alpha  R_{\nu\rho\beta\gamma}\notag\\
	&+\frac\sigma2 \varepsilon ^{\mu\nu\rho} \varepsilon^{\alpha\beta\gamma}
	 u_\alpha  \nabla_\mu R_{\nu\rho\beta\gamma}\notag\\
	 =& \varepsilon ^{\mu\nu\rho} \varepsilon^{\alpha\beta\gamma}
	 \nabla_\rho u_\gamma \left(
	   R_{\mu\nu\beta\lambda}u^\lambda u_\alpha
	 -\frac\sigma2 R_{\mu\nu\beta\alpha} \right)\notag\\
	 =&\varepsilon ^{\mu\nu\rho} \varepsilon^{\alpha\beta\gamma}
	 \nabla_\rho u_\gamma \left(\sigma
	R_{\mu\nu\delta\lambda}	({P_\parallel}^\delta_\beta+{P_\perp}^\delta_\beta)
	{P_\parallel}^\lambda_\alpha
	-\frac\sigma2 R_{\mu\nu\delta\lambda}
	({P_\parallel}^\delta_\beta+{P_\perp}^\delta_\beta)
	({P_\parallel}^\lambda_\alpha+{P_\perp}^\lambda_\alpha) \right)\notag\\
	=	&-\frac\sigma2\varepsilon ^{\mu\nu\rho} \varepsilon^{\alpha\beta\gamma}
	 \nabla_\rho u_\gamma R_{\mu\nu\delta\lambda}{P_\perp}^\delta_\beta
	 {P_\perp}^\lambda_\alpha=0,
\end{align}
where in the first equality the first term vanishes as we have three vectors $\nabla u$ perpendicular to $u$ contracted with an epsilon, and the last term vanishes by the second Bianchi identity. Using the definition of the curvature tensor in terms of covariant derivatives and juggling some indices around we get the second equality. We then insert the identity $g = P_\parallel+P_\perp$ twice and expand. And finally in the last line, we again have three vector perpendicular to $u$ contracted with an epsilon which vanishes. We thus see that this current is in fact identically conserved.

A similar calculation demonstrates under a variation of $u$, $J$ transforms by a total derivative,
\eq{
\delta_u J^\mu = 2 \nabla_\nu \left[
\epsilon^{\mu\nu\rho} \epsilon^{\a\b\gamma}
u_\a \delta u_\b \nabla_\rho u_\gamma
\right],\label{eq:u_vary_J}
}
where we have used the fact that $u\cdot \delta u =0$.

We emphasize that the arguments above require the vector field to be of constant norm. In practical terms, what this means is that we require a nowhere vanishing vector field that we can normalize. As an example, we can consider small fluctuations about a large magnetic field. We will consider this example in detail in the following sections.

\subsection{Conserved topological charge}
\label{subsec:charge}
We now demonstrate that the conserved charge associated with the topological current is the Euler characteristic of the two dimensional surface on which it is calculated. We will first specialize to the case where $u^\mu$, the normalized vector from which we defined $J$ in the previous section, is normal to some surface $\Sigma$ and show the $u\cdot J$ is proportional to $^{(2)}\!R$, the curvature tensor of the submanifold $\Sigma$ and then generalize the result to any surface. 

Assuming that there exists a spatial surface $\Sigma$ orthogonal to $u^\mu$,\footnote{Frobenius' theorem tells us the existence of such a surface requires an integrability condition $\epsilon^{\mu\nu\rho}u_\mu \nabla_\nu u_\rho=0$.} the identity relating the curvature tensor on $\Sigma$ to the full curvature tensor is:
\begin{equation}
{{{^{(2)}\!R}}_{abc}}^d=
	{P_\perp}_a^\alpha{P_\perp}_b^\beta{P_\perp}_c^\gamma{P_\perp}_\delta^d
	 {R_{\alpha\beta\gamma}}^\delta+\sigma\left(K_{ac}{K_b}^d-K_{bc}{K_a}^d\right),
\end{equation}
where $\sigma=\pm1$ is as in the previous section and $K_{ab}$ is the extrinsic curvature associated with the surface $\Sigma$ defined as $K_{ab}=\nabla_a u_b$. Looking at the quantity $u\cdot J$ we have:
\begin{align}
	u\cdot J = \frac1{8\pi}
	\varepsilon ^{\mu\nu\rho} \varepsilon^{\alpha\beta\gamma} u_\mu u_\alpha
		\left( \nabla_\nu u_\beta \nabla_\rho u_\gamma 
		+ \frac\sigma2 R_{\nu\rho\beta\gamma}\right)
			=\frac \sigma{16\pi}
			 \varepsilon ^{\mu\nu\rho} \varepsilon^{\alpha\beta\gamma} u_\mu u_\alpha
				{^{(2)}\!R}_{\nu\rho\beta\gamma}=\frac{\sigma}{8\pi}{}^{(2)}\!R.
\end{align}
Turning this into an integral equation we have:
\begin{equation}
\label{eq:total_charge}
Q_\Sigma=\int\limits_\Sigma \sqrt{^{(2)}\!g\;\,}\; u\cdot J = \frac{\sigma}{8\pi} \int\limits_\Sigma \sqrt{^{(2)}\!g\;}\; ^{(2)}\!R = \sigma \frac{\chi}{2},
\end{equation}
where $Q_\Sigma$ denotes the total charge evaluated on the surface $\Sigma$, $^{(2)}\!g$ is the two dimensional induced metric and $\chi$ is the Euler characteristic of the surface $\Sigma$. 

Of course this was a special case, as by Frobenius theorem it is not always possible to find a surface normal to the vector field $u^\mu$. However, barring topological obstruction, we can find an interpolating function that interpolates between a spacelike hypersurface's normal vector to an arbitrary (smooth and everywhere timelike) unit vector $u_*^\mu$ as a function of time. If we have a family of foliations $\Sigma_t$ with normal vector $n^\mu$ Consider an $L^\infty$ vector of the form
\eq{
u^\mu=\frac{(1-f(t) )n^\mu+f(t)u_*^\mu}{\left\|(1-f(t) )n^\mu+f(t)u_*^\mu\right\|}
,~
f(t) = 
\left\{\begin{array}{cc}
0 &  t \le 0 \\
\cO(t)^3 & 0 \lsim  t\\
1+\cO(1-t)^3 & t \lsim 1 \\
1 & t \ge 1
\end{array}\right.
}
The fact that $n$ and $u_*$ are both unit timelike vectors and the triangle inequality guarantees that $u$ is smooth as well. 
Since the current $J$ is second order in derivatives, it will receive no contribution from the $t^3$ part at $t=0$ and from the $(1-t)^3$ at $t=1$. We can hence argue that this deformation leaves the value of the charge associated with the current unchanged at $t=1$. Also, from the previous argument, it is clear that the charge at $t=0$ is proportional to the Euler characteristic of the two dimensional surface defined by $t=0$. Applying charge conservation, we see that the value of the charge at $t\lsim1$ is also equal to the Euler characteristic. We have hence shown that the value of the charge evaluated on any surface which is cobord to the surface $t=0$ is equal to the Euler characteristic which remains unchanged in time. We note that this does \emph{not} imply that the charge density is equal to the Euler density, and it will of course in general not be. 

Of course, the main assumption here was that the vector field $u$ was smoothly deformable to the constant normal vector $n$. This is true in Minkowski signature thanks to our requirement that $u$ be a smooth vector\footnote{Consider the 2D Euclidean analog where we take $u$ to be the normalized gradient of the height function of a Riemann surface. On foliations where the topology of a constant height section changes the height function must have an extremum, which means that $u$ is not smooth there.}. but does not always hold for  the Euclidean case. We now turn to an example of a topologically non-trivial mapping where we cannot follow the arguments above.

Take $\mathcal M = \mathds R \times S^2$, where we treat the non-compact direction as time. We argued above that with a Minkowski signature and time-like $u^\mu$, the value of the total charge is $-1$.  With a Euclidean signature, however, the value is not fixed. Since the space of unit vectors is a sphere, we can classify different vector fields $u^\mu$ by $\pi_2(S^2)=\mathds Z$. These are of course the winding classes of mapping a sphere to a sphere. Doing the calculation of the total charge on a surface of constant Euclidean time,we find that $Q=n$, where $n$ denotes the winding of $u^\mu$.

We can understand this result as follows. If we look at $\mathds R \times S^2$ as a spherical shell with time flowing radially outwards, then the mapping which we considered above of $u^\mu \sim \d_\tau$, that is the vector field constantly pointing in the Euclidean time direction, would of course correspond to a radially outward pointing vector field. It is clear that this mapping has winding $n=1$ and therefore the charge would be equal to $Q=1$, matching equation \eqref{eq:total_charge}. For general products $\mathds R \times \Sigma$ in $D$ Euclidean dimensions, the vector field at constant $\tau \in \mathds R$ is a map $\Sigma \rightarrow S^{D-1}$ and will have some nontrivial mapping class for $D>1$.

\subsection{Weyl invariance}
\label{subsec:weyl}
 Given the topological nature of the charge corresponding to this new current, it is natural to study how the current transforms under Weyl rescalings of the spacetime metric. Construction of the topological current requires a unit norm vector, and so under a nonstandard Weyl rescaling
\eq{
g_{\mu\nu}\rightarrow \tilde g_{\mu\nu}  = e^{2\Omega} g_{\mu\nu},~u^\mu\rightarrow \tilde u^\mu = e^{-\Omega} u^\mu.
}
This leads to a rather simple transformation property for the current \emph{density} $\sqrt{-g}J^\mu$, 
\eq{
4 \pi\delta_\Omega \left( \sqrt{-g} J^\mu \right) = \sqrt{-g} \nabla_\nu \left(u^\nu \partial^\mu \Omega - u^\mu \partial^\nu \Omega \right). \label{eq:Weyl_trams}
}
The benefit of this simplified transformation of the charge density is easily seen when we construct the charge density two form $(*J)_{\mu\nu} = -\ce_{\mu\nu\rho} J^\rho$, whose pullback to $\Sigma$ is the charge density integrand,
\eq{
\phi_\Sigma^*\left(*J \right) = \sqrt{h_\Sigma} n_\mu J^\mu d\Sigma, ~4\pi \delta_\Omega\left( *J \right)_{\mu\nu}=\nabla_\sigma\ce_{\mu\nu\rho} \left(u^\sigma \partial^\rho \Omega - u^\rho \partial^\sigma \Omega \right).
}
We therefore find that $\delta_\Omega (*J) = d\Lambda$ , and at least for manifolds without boundary, we confirm that the charge density is always Weyl invariant.

With a locally conserved current, it is of course natural to couple it to a $U(1)$ connection. One can check that, without boundary terms,
\eq{
\delta_\Omega \left( \int \sqrt{-g} A_\mu J^\mu \right) =\frac{1}{4\pi} \int \sqrt{-g} F_{\mu\nu} u^\nu \partial^\mu \Omega,
}
and therefore if $F\cdot u =0$, the coupling is Weyl invariant (up to boundary terms.) As we will see when considering this coupling in quantum Hall in section \ref{sec:QH}  and 2+1 superfluids in section \ref{sec:SF}, this constraint is naturally satisfied.

\subsection{Boundary theory}
\label{subsec:anomaly}
On a theory with boundary with normal vector $n_\mu$, $n\cdot J$ is not necessarily zero and hence, to preserve charge conservation, there needs to be a boundary current to compensate for this.  We define:
\begin{equation}
\label{eq:boundary_current}
k^\mu=\frac{1}{4\pi}\varepsilon^{\mu\nu\rho}n_\rho n'^\alpha \nabla_\nu t_\alpha,
\end{equation}
where $t^\mu=\frac1{\sqrt{1+(n\cdot u )^2}}\varepsilon^{\mu\nu\rho}u_\nu n_\rho$ and $n'^\mu=\varepsilon^{\mu\nu\rho} t_\nu u_\rho$. We note that this current has the desired properties of being parallel to the boundary and also the scalar $u\cdot k$ gives the correct contribution such that the total charge of the bulk and boundary currents is equal to the Euler characteristic.  To calculate the divergence of this current, we define the boundary projector: ${P}^\mu_\nu=\delta^\mu_\nu-n^\mu n_\nu$. We see:
\begin{align}
8\pi\,{}^{(2)}\nabla_\mu k^\mu=&  P^\mu_\alpha P^\beta_\mu \nabla_\beta k^\alpha
		= P^\beta_\mu \nabla_\beta k^\mu
		=2 P^\beta_\mu \nabla_\beta \left(\varepsilon^{\mu\nu\rho}n_\rho 
						n'^\alpha \nabla_\nu t_\alpha\right)\\
	=& 2 P^\beta_\mu \varepsilon^{\mu\nu\rho}\left(\nabla_\beta n_\rho \right)
				n'^\alpha (\nabla_\nu t_\alpha)
		+ 2 P^\beta_\mu \varepsilon^{\mu\nu\rho} n_\rho 
				\left(\nabla_\beta n'^\alpha \right)  (\nabla_\nu t_\alpha)
		+ 2 P^\beta_\mu \varepsilon^{\mu\nu\rho} n_\rho 
				n'^\alpha \nabla_\beta \nabla_\nu t_\alpha \notag\\
 	=& 2 \varepsilon^{\beta\nu\rho}n_\nu \left(\nabla_\beta n_\rho \right)
				n'^\alpha n^\gamma \nabla_\gamma t_\alpha
		- 2 \varepsilon^{\beta\nu\rho} n_\rho 
				u_\alpha (\nabla_\beta n'^\alpha) u^\gamma (\nabla_\nu t_\gamma)
		+ \varepsilon^{\beta\nu\rho} n_\rho n'^\alpha
						 R_{\beta\nu\alpha\gamma} t^\gamma\notag\\
	=& 0  - 2 \varepsilon^{\beta\nu\rho} n_\rho 
				n'^\alpha t^\gamma (\nabla_\beta u_\alpha)  (\nabla_\nu u_\gamma)
		+\frac12 \varepsilon^{\rho\beta\nu} \varepsilon^{\lambda\alpha\gamma}
				n_\rho u_\lambda  R_{\beta\nu\alpha\gamma}\notag\\
	=& -  \varepsilon^{\rho\beta\nu} \varepsilon^{\lambda\alpha\gamma}
		n_\rho u_\lambda \big(\nabla_\beta u_\alpha\nabla_\nu u_\gamma
		+\frac12   R_{\beta\nu\alpha\gamma}\big).\notag
\end{align}
Here, in the third line the following manipulations have taken place. On the first term, the $\nu$ index has been projected onto $n$ as no other index in $\epsilon^{\mu\nu\rho}$ can be in the $n$ direction. In the second term, the $\alpha$ index has been projected on to $u$ because $\alpha$ has to be perpendicular to both $t$ and $n'$. And in the last term the identity for the curvature has been used. 

In the fourth line, we note that $t$ can be continued into the bulk in such a way that $n\cdot \nabla t=0$. (One can also say that the derivative in the definition of $k$ is only a boundary derivative and therefore $n\cdot\nabla=0$.) This implies that the first term vanishes. On the second term we have swapped the vectors in the derivative twice. And in the third term (as well as the fifth line, we have used $n'^\alpha t^\gamma-n'^\gamma t^\alpha=\varepsilon^{\lambda\alpha\gamma}u_\lambda$. This coincides with $-8\pi n\cdot J$.

We can also calculate the transformation of the boundary current under a Weyl rescaling:
\eq{
g\rightarrow \tilde g_{ab}  = e^{2\Omega} g_{ab},~u\rightarrow \tilde u_a = e^{\Omega} u_a,~n\rightarrow \tilde n_a = e^{\Omega} n_a,~t\rightarrow \tilde t_a = e^{\Omega} t_a.
}
Under this transformation, the boundary current transforms as:
\begin{align}
\sqrt{-{}^{(2)}\!g\,}\,k^\mu \rightarrow &
	\frac1{4\pi} e^{-\Omega}
	 \sqrt{-{}^{(2)}\!g\,}\, \varepsilon^{\mu\nu\rho}n_\rho n'^\alpha 
		\tilde \nabla_\nu e^{\Omega} t_\alpha
	=\frac{1}{4\pi} \sqrt{-{}^{(2)}\!g\,}\, \varepsilon^{\mu\nu\rho}n_\rho n'^\alpha 
		\tilde \nabla_\nu t_\alpha.
\end{align}
Therefore, the change of the boundary current is:
\begin{align}
\delta_\Omega \left(\sqrt{-{}^{(2)}\!g\,}\,k^\mu\right) 
=& \frac1{4\pi}\sqrt{-{}^{(2)}\!g\,}\, \varepsilon^{\mu\nu\rho}n_\rho n'^\alpha 
		(\delta_\Omega \Gamma_{\nu\alpha}^\beta) t_\beta\notag\\
=& \frac1{4\pi}\sqrt{-{}^{(2)}\!g\,}\,  \varepsilon^{\mu\nu\rho}n_\rho n'^\alpha t_\beta
	\big(2 \delta^\beta_{(\nu} \d^{}_{\alpha)} \Omega - g_{\nu\alpha} \d^\beta \Omega\big)
	\notag\\
=&  \frac1{4\pi}\sqrt{-{}^{(2)}\!g\,}\,  \varepsilon^{\mu\nu\rho}n_\rho 
	\big( n'_\alpha t_\nu  - n'_\nu t_\alpha \big) \d^\alpha \Omega \notag\\
=&  \frac1{4\pi}\sqrt{-{}^{(2)}\!g\,}\,  \varepsilon^{\mu\nu\rho}n_\rho
	\varepsilon_{\lambda \nu \alpha} u^\lambda \d^\alpha \Omega
=  \frac1{4\pi}\sqrt{-{}^{(2)}\!g\,}\, n_\rho (u^\rho \d^\mu - u^\mu \d^\rho) \Omega\notag.
\end{align}
Again, this cancels the variation of \eqref{eq:Weyl_trams}. 

\subsection{Generalizations}
\label{subsec:other_dims}
We can generalize this current to other dimensions as well. From the proof of conservation (eq. \eqref{eq:current_conservation}), it is clear to see that in $d$ dimensional space-time the current defined as:
\begin{equation}
J^\mu_d = \varepsilon^{\mu \mu_1\cdots \mu_{d-1}}\varepsilon^{\nu_1 \cdots \nu_d} u_{\nu_1}
	\D_{\mu_1}u_{\nu_2}\cdots \D_{\mu_{d-1}}u_{\nu_d},
\end{equation}
is conserved in flat space\footnote{This current was discussed in the context of a background isospin field coupling to a Dirac spinor in two, three and four dimensional flat space \cite{Abanov:1999qz}.}. In curved space however, we can only define a conserved current in odd-dimensional spaces, given by:
\begin{multline}
J^\mu_d = \varepsilon^{\mu \mu_1\cdots \mu_{d-1}}\varepsilon^{\nu_1 \cdots \nu_d} u_{\nu_1}
 \left(\D_{\mu_1}u_{\nu_2}\cdots \D_{\mu_{d-1}}u_{\nu_d}
 +\frac\sigma2 \frac{d-1}{2!!}R_{\mu_1\mu_2\nu_2\nu_3} 
 	\D_{\mu_3}u_{\nu_4}\cdots \D_{\mu_{d-1}}u_{\nu_d}\right.\\\left.
 +\left(\tfrac\sigma2\right)^2 \frac{(d-1)(d-3)}{4!!}R_{\mu_1\mu_2\nu_2\nu_3} R_{\mu_3\mu_4\nu_4\nu_5} 
 \D_{\mu_5}u_{\nu_6}\cdots \D_{\mu_{d-1}}u_{\nu_d}\right.\\\left.
 +\left(\tfrac\sigma2\right)^3 \frac{(d-1)(d-3)(d-5)}{6!!}R_{\mu_1\mu_2\nu_2\nu_3} R_{\mu_3\mu_4\nu_4\nu_5} R_{\mu_5\mu_6\nu_6\nu_7} 
 \D_{\mu_7}u_{\nu_8}\cdots \D_{\mu_{d-1}}u_{\nu_d}\right.\\\left.
 +\cdots +\left(\tfrac\sigma2\right)^\frac{d-1}{2}R_{\mu_1\mu_2\nu_2\nu_3}\cdots
 R_{\mu_{d-2}\mu_{d-1}\nu_{d-1}\nu_d}\right).
\end{multline}
Again, we can demonstrate, following the arguments of section \ref{subsec:charge} that if there is a codimension one hypersurface $\Sigma$ perpendicular to the vector field $u^\mu$, then:
\begin{equation}
u\cdot J_d = \left(\tfrac\sigma2\right)^\frac{d-1}{2} \varepsilon^{\mu \mu_1\cdots \mu_{d-1}}\varepsilon^{\nu_1 \cdots \nu_d} u_\mu u_{\nu_1} \,^{(d-1)}\!R_{\mu_1\mu_2\nu_2\nu_3}\cdots
 \,^{(d-1)}\!R_{\mu_{d-2}\mu_{d-1}\nu_{d-1}\nu_d},
\end{equation}
where again $\,^{(d-1)}\!R_{\mu\nu\rho\sigma}$ denotes the Riemann curvature tensor of the $d-1$-dimensional submanifold $\Sigma$. We recognize this as the $(d-1)$-dimensional Euler density. Integrating over $\Sigma$ we get:
\begin{equation}
\int\limits_\Sigma \sqrt{^{(d-1)}\!g\;\,}\; u\cdot J_d =(4\pi\sigma)^\frac{d-1}{2}
\left(\tfrac{d-1}2\right)!\, \chi(\Sigma).
\end{equation}
This justifies the fact that the conserved current can only be constructed in odd dimensions, as the Euler characteristic is only defined in even dimensions. Note that the three dimensional current defined previously is $J^\mu=\frac1{8\pi}J^\mu_3$.

It is also straight-forward to rewrite this current in the first order formalism. Taking $e^a_\mu$ and ${\omega^a_\mu}_b$ to be the tetrad and the connection 1-form respectively, the 3-dimensional current in first order formalism would most naturally be written:
\begin{equation}
\label{eq:firstorderformalism}
	J=*\Theta, \;\; \Theta = \epsilon^{abc} u_a \left(D u_b \wedge D u_c - R_{bc} \right),
\end{equation}
where $u_a=e^\mu_a u_\mu$ and $D u_a = du_a - \omega ^b_a u_b$ is the covariant exterior derivative of $u$. In this language, the statement of current conservation is $d\Theta=0$. The generalization of this form of the current to higher dimensions follows exactly as above. The advantage of this notation, however, is that it would also be straightforward to generalize to some other gauge groups. In this case $u_a$ would be a representation and $\omega$ and $R$ would have to be replaced by $A$ and $F$, the connection and curvature of the gauge group respectively. An example of such structures is global angular forms \cite{bott1982lw}, used in algebraic topology and occasionally in high energy contexts (for instance \cite{Freed:1998tg}).
 
\section{Quantum Hall}
\label{sec:QH}
In this section we review the and expand on the results of \cite{Golkar:2014}. As discussed in \cite{Golkar:2014}, the inclusion of the topological current is required to match the correct degeneracy of the Landau levels similar to the Wen-Zee term in the non-relativistic case. Our goal is to construct an effective field theory to describe Lorentz invariant quantum Hall systems after all heavy degrees of freedom have been integrated out, and we are left with a generating functional
\eq{
Z[A_\mu, g_{\mu\nu}] = \exp\left( iS_{eff}[A_\mu, g_{\mu\nu}]\right)
}
which we require must be invariant under both $U(1)$ and diffeomorphism invariant. Since we want the effective theory to be applicable to quantum Hall systems, $F^2>0$, we will work in a derivative expansion about a background with a constant magnetic field, and do not  require the action to be smooth as $B \rightarrow 0$, where  the Landau level spacing would vanish.

While we have discussed in section \ref{sec:current} the generalization of the Euler current to manifolds with boundary, in what follows we will assume that the theory is defined on a manifold without boundary.  The inclusion of the boundary and its implications for the existence of light modes  is the subject of a forthcoming study.

\subsection{Effective action and power counting}
\label{subsec:QH-EFT}
We will work in natural units where $\hbar = v_F = e = 1$. Our power counting scheme is designed  so that the external magnetic field $B$ is of $\cO(1)$ and hence $F_{\mu\nu}=\cO(1)$. This implies that the gauge field $A_\mu=\cO(p^{-1})$. The metric is assumed to have perturbations about the flat background which are of order one, $g_{\mu\nu}=\cO(1)$ such that the Riemann curvature would be of order $p^2$. 

In order to include the current $J^\mu$ discussed in the previous section, we need a normalized time-like vector field. In the problem at hand, since we are considering fluctuations of the background electromagnetic field about a large magnetic field $B$, we would expect $F_{\mu\nu}F^{\mu\nu}$ to be always greater than one. We can hence construct:
\begin{equation}
u^\mu=\frac{1}{2b}\varepsilon^{\mu\nu\lambda} F_{\nu\lambda}, 
b=(\frac12 F_{\mu\nu}F^{\mu\nu})^\frac12,
\end{equation}
which has the desired properties. Physically, $u^\mu$ is the local frame in which the field is purely magnetic, and $b$ is the magnetic field in that frame.  Note that the Bianchi identity $dF=0$ turns into a  conservation equation $\nabla_\mu (b u^\mu)=0$. 
 Both $u^\mu$ and $b$ are $\cO(1)$ in our power counting scheme and we will use them as building blocks of the gauge invariant terms in the action, as power counting is a simple derivative counting on $b$ and $u^\mu$. With this power counting, the Chern-Simons term is $\cO(p^{-1}),$ and the only $\cO(1)$ term we can write down is a scalar function of the magnetic field $\e(b)$.
The action to order $\cO(p^0)$ describes a perfect fluid with stress tensor given by:
\eq{
T^{\mu\nu}=(\epsilon+P)u^\mu u^\nu + P g^{\mu\nu}, 
P=b \epsilon'(b) - \epsilon(b).
}
The term $A_\mu J^\mu$ is $\cO(p)$.
There are two other terms that one can write down at $\cO(p)$:
\begin{equation}
	f_1(b)\varepsilon^{\mu\nu\lambda} u_\mu \d_\nu u_\lambda, 
	f_2(b)u^\mu\d_\mu b.
\end{equation}
The second term, however, is seen to be a total derivative if we define $f_2(b)=b f'_3(b)$, that is $f_2(b)u^\mu\d_\mu b=\nabla_\mu (b u^\mu f_3(b))$. Hence, the full Lagrangian including all possible terms to order $\cO(p)$ is:
\begin{equation}\label{L-eff}
  \mathcal L = \frac\nu{4\pi} \varepsilon^{\mu\nu\lambda}
  A_\mu\d_\nu A_\lambda - \varepsilon(b) 
  + f(b) \epsilon^{\mu\nu\lambda}   u_\mu\d_\nu u_\lambda   + \kappa A_\mu J^\mu.
\end{equation}
It is interesting to note that when $\e(b)\sim b^{3/2}$ and
$f(b) \sim b$, the action is fully Weyl invariant. This would
be the case if the microscopic theory underlying the quantum Hall state is a conformal field theory, in which case all the non-universal functions appearing in the effective action up to first order in derivatives are fixed. In all that follows, the results for a Weyl invariant system can be easily read off by making the above substitutions.

\subsection{Relativistic shift}
\label{subsec:QH-shift}

If we calculate the charge density of the theory by taking a variation with respect to $A_0$, we see that the only contributions come from the Chern-Simons term and the new topological current:
\begin{equation}
	Q=\int_\Sigma \left(\frac{\nu}{4\pi} n_\mu \ce^{\mu\nu\rho}F_{\nu\rho} +  \kappa n_\mu J^\mu \right) = \nu N_\phi + \frac \kappa 2 \chi.
	\label{eq:QH_shift_defined}
\end{equation}
Where $N_\phi=\frac{1}{2\pi}\int_\Sigma F$ is the total magnetic flux quanta through the spatial manifold, $\chi=2(1-g)$ is the Euler character, and $n^\mu$ is the normal vector to our surface $\Sigma$. This of course, is the relativistic analogue of the shift. For comparison, in the non-relativistic case, the shift $\mathcal S$ is defined via $Q=\nu (N_\phi + \mathcal S)$, i.e., $\nu \mathcal S$ is the offset between the total charge and the fraction of the total flux. However, since in a relativistic theory, we can have a finite offset even at zero filling fraction, we have defined the offset (that is $\kappa \chi/2$) to be independent of $\nu$, so that it remains finite when $\nu=0$.

In order to match the coefficient $\kappa$ to a microscopic description, we look at the degeneracy of the Landau levels on a sphere. The situation is similar to the non-relativistic case \cite{Greiter:2011}, with two minor differences. Due to the nontrivial spin connection on the sphere, we can not simply square the Dirac Hamiltonian and use the Schr\"odinger equation degeneracy. For instance, the lowest Dirac Landau level on a sphere with $N_\phi$ magnetic flux is the same as the lowest Schr\"odinger Landau level with $N_\phi^{\text{NR}}=N_\phi-1$ flux quanta, and therefore has degeneracy $N_\phi = N_\phi^{\text{NR}}+1$.  The degeneracy of the $n^\mathrm{th}$ Landau level is similarly $N_\phi+2n $. Second, the energy spectrum is offset, such that the ground state has zero energy, which is of course the source of the offset of $\frac12$ in the filling fractions.

For an integer quantum Hall state with $\nu= N_f(n+\frac12)$, where $N_f$ is the total ``flavor" degeneracy of the Landau levels ($N_f=4$ for graphene), the total charge
can be found by summing up all charges of the filled Landau levels. Defining the half-full, particle-hole symmetric zero energy level to carry zero charge gives $\kappa = N_f n (n+1)$. Note that $\kappa=0$ for the $\nu=\pm2$ states in graphene, corresponding to full and empty zero energy Landau levels.

A similar counting holds for fractionally filled Landau levels. For illustration, we look at graphene with complete  $SU(4)$ flavor symmetry breaking. Consider the state with $0<\nu<1$. Of the four zero energy Landau levels, two are full, one is partially filled and the last is empty (the sequential complete filling of each of these characterized by $\nu=-1,0,1,2$ respectively). To proceed further, we need to know the charge offset at fractional filling. Comparing again with the non-relativistic case where $Q=\nu (N'_\phi + \mathcal S_{\text{NR}})$, we see that $\kappa=\nu (\mathcal S_{\text{NR}}-1)$. For instance, in the case of Laughlin fractions $\nu=\frac1m$, $\kappa=\frac{m-1}{m}$.

\subsection{Discrete symmetries}
\label{subsec:QH-CPT}
We now discuss how the effective theory transforms under $C$, $P$, and $T$. We start with the standard $CPT$ transformations of three dimensional abelian gauge theories \cite{Deser:1981wh},
\eqn{
C&:&A_\mu \rightarrow - A_\mu, \nonumber \\
P&:&x^1 \rightarrow -x^1,~A_0 \rightarrow A_0, ~A_1 \rightarrow -A_1, ~A_2 \rightarrow A_2, \nonumber \\
T&:&x^0 \rightarrow -x^0,~A_0 \rightarrow A_0, ~A_i \rightarrow -A_i.
}
All three individual symmetries are broken by a background magnetic field, and a nonzero chemical potential breaks $C$. The combination $PT$ is preserved by both the magnetic field and chemical potential. One finds that all terms in our effective action \eqref{L-eff} are invariant under $PT$. Turning off a chemical potential, we can  classify all terms with respect to $CP$ and $CT$. Under both of these, the $\nu,~\kappa$ and $u\wedge du$ terms are all odd and therefore nonzero values at zero chemical potential indicate a spontaneous breaking of $CP$ and $CT$. As argued in \ref{subsec:QH-shift}, studying the shift of the $\nu=0$ integer quantum Hall state of graphene we find $\kappa=0$, consistent with unbroken symmetries. On the other hand, it is easy to construct a multi-flavor Moore-Read state \cite{Moore:1991ks} at $\nu=0$ which breaks both $CP$ and $CT$.

\subsection{Momentum density and the microscopic theory}
\label{subsec:QH-micro}

As a first example of what we can learn from our effective theory, we  compute the momentum density
$T^{0i}$ in the background of a static inhomogeneous magnetic field $B=b(x,y)$. We turn on a perturbation in the $g_{0i}$ component of the metric tensor and read out the momentum density from the action: $\delta S =
\int\!d^3x T^{0i}g_{0i}$.  We find
\begin{equation}
	\label{eq:mom_density}
  T^{0i} = - \epsilon^{ij}\d_j \left(\frac\kappa{8\pi}b+f(b)\right) .
\end{equation}
Note that if we use $L=r\times p$ to construct an angular momentum density we trivially get zero.
For
the FQH states on the zeroth Landau level with negligible mixing with
other Landau levels, the function $f(b)$ is completely determined by
the topological coefficients $\nu$ and $\kappa$.  This comes from a
holomorphic constraint relating the momentum density and the particle
density.

For concreteness, let us work with the standard Dirac representation of the Clifford algebra,
\begin{equation}
  \gamma^0 = \sigma_3, \quad \gamma^1 = i\sigma_1, \quad
  \gamma^2 = i\sigma_2,
\end{equation}
for which the free Dirac Hamiltonian at zero chemical potential has the form
\begin{equation}
  H = -i \gamma^0\gamma^i D_i= 2\begin{pmatrix}
      0 & D \\ -\bar D & 0 \end{pmatrix}.
\end{equation}
Here $D\equiv D_z= \nabla_z-i A_z$ and its conjugate $\bar D\equiv D_{\bar z}$, and we use complex coordinates $z = x+iy$, $\bar z = x-iy$. States in the $n=0$  level are simply zero energy eigenstates of the Hamiltonian, $\psi = (\varphi,0)^T$, with $\varphi$ satisfying
the holomorphic constraint $\bar D \varphi =\bar D \varphi^* = 0$.  
Now let us look at the stress-energy tensor,
\begin{equation}\label{Tmunu}
  T^{\mu\nu} = -\frac i4 \overline\psi \gamma^{(\mu} \Dlr{}^{\nu)} \psi,
\end{equation}
again assuming a static inhomogeneous magnetic field and no
electric field.  For the $0i$ components we can ignore time
derivatives as $A_0=0$ and the lowest Landau level has zero energy.
We see that
\begin{equation}
  T^{0i} = -\frac i4 \varphi^* \Dlr{}^i \varphi,
\end{equation}
which after changing to complex coordinates and using the holomorphic constraints becomes
\begin{equation}
  T^{0i} = -\frac14 \epsilon^{ij}\d_j n,
\end{equation}
where $n=\varphi^*\varphi$ is the particle number density on the
lowest Landau level.  Comparing  to \eqref{eq:mom_density} we find
that for FQH states in the zeroth Landau level,
\begin{equation}
  f(b) = \frac1{8\pi}(\nu-\kappa) b.
\end{equation}
The calculation above neglects possible mixing between Landau levels,
as well as corrections to the stress-energy
tensor~(\ref{Tmunu}) due to interactions. Both effects are small when
the interaction energy scale is much smaller than the distance between
Landau levels $\sqrt B$. In particular, we know that turning on Coulomb interactions with a strength $e^2$ will partially split the degeneracy of the Landau level and correct the stress tensor, but the result will be
\eq{
T^{0i} = -\frac{\nu}{8\pi}\epsilon^{ij} \partial_j b + e^2 \times T_{int.},
}
which means that we can trust our result so long as the Coulomb interaction is weak\footnote{Note, of course, that even at small $e$ degeneracy splitting is a nonperturbative problem.}.

\subsection{Response functions}
\label{subsec:QH-response}

We now compute different response functions of the relativistic quantum Hall states to external fields. We can compute the conductivity directly from Ohm's law, $j_i = \sigma_{ij}E^i$, by looking at the linearized current in flat space. We find the conductivity matrix at nonzero
frequencies and wavenumbers,
\begin{equation}
\sigma_{xx}(\omega, \mathbf k) = -i\omega \frac{\ce'(B)}{B}+\cO(k^3)
,~
  \sigma_{xy}(\omega,\mathbf k) = \frac\nu{2\pi} +
   \left( \frac\kappa{4\pi} + \frac{2f(B)}B \right) \frac{\omega^2}B
   - \frac{f'(B)}B \frac{\mathbf{k}^2}B+\cO(k^4)\,.
\end{equation}
At lowest Landau level, the Hall conductivity simplifies to
\begin{equation}\label{Kohn}
  \sigma_{xy}(\omega,\mathbf{k}) = \frac\nu{2\pi} + \frac\nu\pi \frac{\omega^2}B
    + \frac{\kappa-\nu}{8\pi}\frac{\mathbf{k}^2}B+\cO(k^4,e^2).
\end{equation}
In Galilean invariant systems, the frequency dependence of the
conductivity matrix is completely determined, at $q=0$, by Kohn's
theorem~\cite{Kohn:1961zz}.  In relativistic systems Kohn's theorem no
longer applies.  Nevertheless, Eq.~\eqref{Kohn} implies that at least in the zeroth Landau level the $\omega^2$ correction is completely
fixed by the filling fraction.

We next compute the Hall viscosity, defined as parity odd response to uniform shear metric perturbations. Specifically in 2+1 dimensional theories with parity broken,
\eq{
\C{T_{xx}(-\omega,0)T_{xy}(\omega,0)}=-i\eta_H\omega+\cO(\omega^2)
.}
Turning on only spatially homogeneous perturbations of the
spatial metric, $g_{ij}=\delta_{ij}+h_{ij}(t)$, one finds the only term that contributes is the Euler current coupling, giving
\begin{equation}
 \kappa \int \!d^3x \,\sqrt{-g} A_\mu J^\mu = 
  	-\frac{\kappa B}{32 \pi} 
  	\int \!d^3x \, \epsilon^{jk} h_{ij} \d_t h_{ik},
\end{equation}
yielding
\begin{equation}
  \eta_H = \frac{\kappa B}{8 \pi}\,.
\end{equation}
The relationship between the Hall viscosity and $\kappa$ is identical to the nonrelativistic result $\eta_H = n \mathcal{S}/4$ \cite{ReadRezayi:2010} with the substitution $\mathcal{S} \rightarrow \mathcal{S}_{\text{NR}}-1$ for graphene states with $0<\nu<1$.
Note that the Hall viscosity depends only on the topological number
$\kappa$. This is expected since $\eta_H$ can be determined by adiabatic
transport and hence should not depend on non-universal functions like
$f(b)$. Note that we can  calculate angular momentum density from \eqref{eq:mom_density} and find that it vanishes, so our effective action does not reproduce the relationship between Hall viscosity and orbital angular momentum density \cite{Read:2008rn,ReadRezayi:2010}.

Next we look at the components of the stress tensor when one turns
on a static, spatially inhomogeneous, electric field.  The result is
\begin{equation}
  T_{ij} = P\delta_{ij} + \frac{\kappa}{8\pi} (\d_i E_j + \d_j E_i) 
  	- \left(\frac{\kappa}{4\pi}+f'(B)\right) \delta_{ij} 
  \bm{\nabla}\cdot {\bf E}. 
\end{equation}
Rewriting this in terms of the drift velocity
$v^i=\epsilon^{ij}E_j/B$ and the shear rate
$V_{ij}=\frac12(\d_iv_j+\d_jv_i -\delta_{ij}\d\cdot v)$, 
\begin{equation}
 T_{ij} = P\delta_{ij} - \eta_H (\epsilon_{ik}V_{kj}+\epsilon_{jk}V_{ki}) + \delta_{ij} (\eta_H+Bf'(B)) \bm{\nabla}\times {\bf v}.
\end{equation}
The traceless part of the stress tensor reflects the nonzero Hall
viscosity of the quantum Hall
fluid~\cite{Avron:1995fg,Avron:1997,Read:2008rn}.

Finally, we look at the response of the system to a gradient in the gravitational potential. Turning on a small contribution to $g_{tt}=-1+h_{tt}$, we find:

\begin{equation}
\langle T^{tx}\rangle= \left(\frac{ \kappa}{4\pi} B_0   + f(B_0) \right) \d_y h_{tt}.
\end{equation} 

\section{Superfluids}
\label{sec:SF}

In this section we consider the $2+1$ dimensional superfluid and look at the contributions of the new terms to parity odd transport.  The overall structure of the Lagrangian and the power counting as well as the discrete symmetries here mirror closely those in the quantum Hall case. For the sake of completeness, we repeat the necessary arguments in the following sections.  

\subsection{Effective action and power counting}
\label{subsec:SF-EFT}

The standard formulation of the effective theory of a
superfluid~\cite{Son:2002} is in terms of a single Goldstone mode
$\psi$ with a shift symmetry which is the phase of the superfluid
condensate.  In three dimensions we can use an alternative dual
description in terms of a $U(1)$ gauge field, $f=da\sim *d\psi$.  The
particle number current is then 
\begin{equation}
  j = \frac 1{2\pi} *\!\!f.
\end{equation}
The fluid conservation equation $d*j=0$ is simply the Bianchi identity for $f$. The field $f=da$ should not
be confused with the external electromagnetic field $F=dA$, which is
nondynamical and coupled via $q A\cdot j$. We will be
interested in expanding about backgrounds of the form: 
\eq{
  j^\mu = n_0 (\d_t)^\mu,~A = \mu_A
  dt,~g_{\mu\nu}=\eta_{\mu\nu}, } 
which correspond to configurations
close to a stationary fluid in a flat background at constant finite
chemical potential.

Similar to the previous section, power counting in momenta $p$, we find that $f\sim \mathcal O(1)$, and therefore $a\sim \mathcal O(p^{-1}).$ The only difference now is that $A\sim \mathcal O(1).$ Now, under the reasonable assumption that we do not want a parity-breaking and mass-generating Chern-Simons term $a\wedge da$,\footnote{The massless excitations of this theory are indeed the Goldstone modes of spontaneously breaking the superfluid $U(1)$ symmetry generated by the current $\epsilon^{\mu\nu\rho} f_{\nu\rho}$  \cite{Kovner:1990pz}. Hence any term that would gap the system, such as $a\wedge da$, would be prohibited by this symmetry.} which would gap out the Goldstone boson, the most general action at lowest order $\cO(p)^0$ is
\eq{
S_0 = \int\!d^3x\,\sqrt{-g} \left[
-\ce(n)+ \frac q{4\pi} \varepsilon^{\mu\nu\rho}A_\mu f_{\nu\rho}
 \right],
}
where $n^2 = -j^\mu j_\mu=+f^2/8\pi^2$.

Before including higher derivative terms, let us study this effective theory in some detail. The first variation of this action gives the lowest order equations of motion
\begin{equation}\label{SF-eom}
\frac{1}{4\pi^2}  \D_\mu \Bigl(\frac{\ce'(n)}n f^{\mu\nu}\Bigr)
  + \frac q{4\pi} \varepsilon^{\nu\rho\lambda}F_{\rho\lambda}=0,
\end{equation}
or in form notation,
\begin{equation}
\label{eq:zeroorderEOMform}
\frac{1}{4\pi^2} d*\!\left(
\frac{\ce'}{n}f
\right)+\frac{q}{2\pi}F=0.
\end{equation}
We can also construct the zeroth order stress tensor and electromagnetic current,
\eq{
T_{\mu \nu} = -\ce g_{\mu \nu}
 +\frac{1}{4\pi^2} \ce' f_{\mu \rho}f_\nu{}^\rho, \quad
  j_{EM}^\mu = \frac{q}{4\pi} \varepsilon^{\mu \nu \rho} f_{\nu \rho} = q j^\mu.
}
The current is conserved thanks to $d^2=0$.  If  
we decompose it into a number density and four-velocity,
\eq{
  j^\mu = n u^\mu,\qquad u^\mu = \frac{j^\mu}{\sqrt{-j^2}}
}
then the stress energy tensor will have the form
\begin{equation}
  T^{\mu\nu} = n\ce' u^\mu u^\nu + (n\ce' -\ce)g^{\mu\nu}
\end{equation}
If we now identify $\varepsilon(n)$ with the energy density, as a function
of the particle number density, and 
recall the relationship between pressure and energy density in
zero-temperature thermodynamics: $p=n\epsilon'-\epsilon$, then we see 
that the stress tensor can be written in the familiar ideal fluid form
$T^{\mu\nu} =
(\varepsilon+p)u^\mu u^\nu+p g^{\mu\nu}$. We can also use the thermodynamic identities to identify the chemical potential $\mu = \ce'(n).$
Expanding the action to quadratic order, we can also identify the 
superfluid sound 
speed $c_s^2 = dp/d\rho = n\ce''/\ce'.$

Let us now work to one higher order in derivatives. By the power counting above, $u^\mu$ and $n$ are $\mathcal O(p^0)$ and $J$ the topological current is  $\mathcal O(p^2)$. The lowest order coupling of the superfluid to the topological current, $a\cdot J$ is of order $\mathcal O(p)$. The arguments in section \ref{subsec:QH-EFT} can be repeated to show that the only other term up to redefinitions we need to consider is $\zeta(n) u \wedge du$.\footnote{There is one additional possible term $h(n) f_{\mu\nu}F^{\mu\nu}\propto n h(n)u \wedge F $, but it can be removed by a field redefinition of $a$ along with a shift of $\zeta$.} We therefore need to study the effect of two $\cO(p)$ terms in the action,
\eq{
S_1 = \int \sqrt{-g} \left[ 
\zeta(n) \epsilon^{\mu\nu\rho}u_\mu\d_\nu u_\rho
+ \kappa a_\mu J^\mu
\right],
}
on the transport properties of the system. Similar to the quantum Hall case discussed above, the entire action $S=S_0+S_1$ would be Weyl invariant if we take $\e(n)\sim n^{3/2}$ and $\zeta(n) \sim n$.

We note that the term $a_\mu J^\mu$ can be considered a Wess-Zumino term for a fluid. In a recent paper, all possible Wess-Zumino terms of multiple systems including the  $2+1$ dimensional superfluid were systematically studied \cite{Nicolis:2014}. However, the coupling to the topological current here was not found. We believe this is due to the fact that while the current $J$ is a local function of $d\psi$ and its derivatives the coupling to $a$ looks nonlocal when the superfluid is described using the Goldstone mode.

\subsection{Discrete symmetries}
\label{subsec:SF-CPT}

The CPT properties of the dual gauge field in our superfluid description are similar to those of the previous section. Explicitly, we have:
\eqn{
C&:&A_\mu \rightarrow - A_\mu,a_\mu \rightarrow - a_\mu,
 \nonumber \\
P&:&x^1 \rightarrow -x^1,~A_{0,2} \rightarrow A_{0,2}, ~A_1 \rightarrow -A_1, ~a_{0,2} \rightarrow -a_{0,2}, ~a_1 \rightarrow a_1,  \nonumber \\
T&:&x^0 \rightarrow -x^0,~A_0 \rightarrow A_0, ~A_i \rightarrow -A_i,~a_0 \rightarrow -a_0, ~a_i \rightarrow -a_i,
}
where we have derived the transformation properties of the dual gauge field, $a_\mu$ from the transformations of the goldstone mode in the microscopic theory. It is straight-forward to see using these transformations that the zeroth order action $S_0$, is $C$, $P$ and $T$ even. In turn the first order action $S_1$ is $C$ even but $P$ and $T$ odd. 

\subsection{The superfluid shift}
\label{subsec:shift}
We now show that the effective theory of the superfluid exhibits a
nonzero shift: when put on a sphere with nonzero $\kappa$, our effective field theory needs a nonzero magnetic flux through the sphere to have a smooth ground state. We may rearrange our action as
\begin{equation}
  S =\int\!d^3x\, \sqrt{-g}\,\left[
  \cL[f]+
   a_\mu \left(
    \frac q{4\pi}\varepsilon^{\mu\nu\lambda} F_{\nu\lambda}
    + \kappa J^\mu \right) \right],
\end{equation}
where $\cL_f$ is a gauge-invariant Lagrangian density, depending only on $f_{\mu\nu}$ and its derivatives.  The field equation for $a$ is 
\eq{
\nabla_\mu \cD^{\mu\nu}
= \frac{q}{4\pi} \epsilon^{\nu\rho\lambda}F_{\rho\lambda} + \kappa J^\nu,\label{eq:gen_SF_eom}
}
where in analogy with electromagnetism in medium we define the displacement tensor $\cD^{\mu\nu}$,
\eq{
\cD^{\mu\nu} = 2 \frac{\delta \cL[f]}{\delta f_{\mu\nu}}+2 \kappa \frac{\delta u_\beta}{\delta f_{\mu\nu}} \left(\epsilon^{\lambda\sigma\rho}f_{\lambda\sigma} \epsilon^{\a\b\gamma} u_\a \nabla_\rho u_\gamma \right).
}
which is still antisymmetric and a function of $f$ and its derivatives. It is clear that if we think only in terms of $\cD$, the only change, compared to the zeroth order equation of motion \eqref{SF-eom}, is that now there is an extra source term. 

Note that here and in what follows we assume that there are no superfluid vortices, which would act as delta function electric sources for $\nabla_\mu \cD^{\mu\nu}$ in \eqref{eq:gen_SF_eom}. This can be seen by noting that vortices, which are infinitely massive in the limit where the electron binding energy gap goes to infinity, will couple to the effective theory via a worldline source $\Delta  S = \int_\gamma a$, where $\gamma$ is the worldline of the vortex. This will add a delta function source to the right hand side of \eqref{eq:gen_SF_eom}. However in our effective theory we will then find that near the delta function, \eqref{eq:gen_SF_eom} implies that $\cD^2$ and therefore $f^2$ will change sign and our effective field theory will break down. We therefore can not have any vortices and trust our effective theory (unless the source is cancelled by a singular magnetic flux tube). 

If we have a smooth $\cD$ and therefore a smooth timelike $u$ on a closed manifold, integrating the $\nu=0$ component of \eqref{eq:gen_SF_eom} gives
\begin{equation}
  q N_\phi + \kappa \frac\chi 2 = 0.
  \label{eq:SF_shift_defined}
\end{equation}
This is analogous to the definition of the shift for the quantum Hall case in  \eqref{eq:QH_shift_defined}. There, the shift represents a mismatch between the total magnetic flux and the total charge on a closed surface. Here, the equations of motion require the total charge associated to $a_\mu$ to be zero and hence we derive a relationship between the total flux and the Euler characteristic of the surface, which as we will see below, if not satisfied signifies the presence of singularities in the field configurations. To be precise, we define the superfluid shift to be the net flux required to have no vortices when the superfluid is placed on a sphere:
\eq{
\cS =-\frac{\kappa}{q}.
}

As an example, consider the $p_x+ip_y$ superfluid, whose order parameter defines a vector in the tangent space of the \emph{spatial} manifold which transforms under the electromagnetic $U(1)$, on a sphere. First we look at the case where there is no magnetic field present.  Since we cannot comb a sphere with a vector field,  it is clear that there will be singular points in any configuration, that is there will be superfluid vortices present. However, in the presence of a net magnetic field piercing the sphere, the combing problem is more subtle. We know that in this case the gauge field $A_\mu$ must be defined in patches with transition functions (which depend on the magnetic flux) that tell us about the patching procedure. Now, since the order parameter transforms covariantly under gauge transformations, it too must be defined in patches with the same transition functions. With this procedure, it can be shown that it \emph{is} possible to comb the sphere if there is an appropriate amount of magnetic flux. This argument can be generalized to any Riemann surface and the constraint for having a globally well defined covariant vector field is $q N_\phi -\chi=0$. Comparing to \eqref{eq:SF_shift_defined}, we can read off $\kappa=-2$. Setting the charge $q=-2$ for the order parameter of $p+ip$ we find that the shift is $\cS=-1$. 

A quick way of deriving this result is by looking at the condition of having a globally covariantly constant vector field, where the covariant derivative takes into account both the geometric and the electromagnetic connection, respectively $\omega$ and $A$:
\begin{equation}
 \nabla_\mu v^a = \d_\mu v^a +{\omega_\mu}^{ab} v^b - i q A_\mu v^a,
\end{equation}
where $a$, $b$ are Lorentz indices which are raised and lowered using $\delta^{ab}$ since the discussion is in Euclidean signature.
Defining $\omega_\mu={\omega_\mu}^{ab} \epsilon^{ab}$, $v=v_x+iv_y$ and $\bar v=v_x-iv_y$, we find:
\begin{equation}
 \nabla_\mu v = \d_\mu v + i \Big(\frac12 \omega_\mu - q A_\mu \Big) v,
\;\;\;\;
  \nabla_\mu \bar v 
  = \d_\mu \bar v - i \Big(\frac12 \omega_\mu + q A_\mu \Big) \bar v,
\end{equation}
We can now derive the above relationship between $N_\phi$ and $\chi$ by requiring the total connection to vanish. However, note that this can only be done either for $v$ or for $\bar v$ and not for both. Physically, this means that the $p+ip$ and $p-ip$ superfluids cannot be simultaneously made vortex-free on the sphere. In other words, the sign of the shift depends on the helicity of the order parameter. This discussion resembles the arguments of \cite{Read:1999fn}, which derive the same result by looking at the Bogoluibov-de Gennes equations in the microscopic theory.

\subsection{Transport Coefficients}
\label{subsec:SF-trans}
We now look at the transport coefficients of the system. The full
response of the system to background perturbations can be seen in
appendix \ref{App:SF}. Here, we summarize some key parity odd
transports.

Starting with Hall conductivity we have:
\begin{equation}
\sigma_{xy}(\omega,\mathbf k) =\frac{2n_0 \zeta'(n_0) \mathbf k^2- (4 \zeta(n_0)+\kappa n_0)\omega^2}{2\mu^2(c_s^2\mathbf k^2 - \omega^2)}
\end{equation}
In particular, we can read off the zero frequency  Hall conductivity as:
\begin{equation}
	\sigma_{xy}=\frac{\kappa q^2 n_0}{2 \mu^2}+\frac{2q^2\zeta(n_0)}{\mu^2}.
\end{equation}

We next look at the Hall viscosity by turning on a metric perturbation. We calculate:
\begin{equation}
\langle T_{xx}(-\omega,-\mathbf k) T_{xy}(\omega,\mathbf k)\rangle =
i\omega
 \frac{\kappa n_0}{4 } \frac{\omega^2- 2 c_s^2 k_y^2 }{c_s^2 \mathbf k^2-\omega^2}.
\end{equation}
From which we see:
\begin{equation}
	\eta_H=\frac{\kappa n_0}{4},
\end{equation}
which again matches exactly the value derived in the previous section. Lastly, we calculate the momentum current generated by a gravitation potential gradient via a small perturbation $g_{tt}=-1+\delta h_{tt}(y)$,
\begin{equation}
\langle T^{tx}\rangle=  -n_0\frac{\kappa}{4 c_s^2} \d_y h_{tt}. 
\end{equation} 
In a finite sample, there will be contributions to the momentum flow coming from the boundary terms. This effect as well as the relationship to the thermal Hall conductivity will be analyzed in a forthcoming paper.

\section{Conclusions}
\label{sec:conclusions}

We have constructed a new current in odd dimensions, whose charge is the Euler characteristic of the codimension one hypersurface on which it is calculated. We showed that it is identically conserved but its construction requires the existence of a vector field of unit norm.

We looked at the effects of this term in two scenarios where a normalized vector field is present: the quantum Hall system and the three dimensional relativistic superfluid. In the quantum Hall case, we showed that the topological current is necessary to correctly describe the offset between the magnetic flux and the total charge, i.e., the shift. We also showed that the shift defined in such a way satisfies the same relationship with the Hall viscosity as in the non-relativistic case.

For the superfluid, the topological current describes a parity odd Wess-Zumino term which allows for nonzero Hall viscosity and is only present (as a local gauge invariant term) in the gauge field description. This term also describes the superfluid shift: the amount of magnetic flux needed to pierce a sphere to allow the superfluid to have a vortex-free ground state. In this case, as well as the quantum Hall, the shift only depends on the topology of the spatial manifold of our theory.

It would be interesting to look at the effects of this current in other systems as well. In this paper we only analyzed the properties of systems in three dimensions. However, the current can be written in any odd dimensional system, where one can define a nowhere vanishing vector field. This of course would have many applications where the emergent properties of the low energy system rely on the presence of a background vector field (similar to the quantum Hall case) or when there is a vector-like dynamical degree of freedom, fluctuating about a non-zero background (similar to the superfluid case). 

Another direction of generalization of these arguments is to extend the discussion to systems in backgrounds with torsion, that is, in the first order formalism. This would allow us to analyze the effects of this current on the spin current by looking at the response of the system to fluctuations in torsion and is the subject of a forthcoming paper.

\section*{Acknowledgments}
It is a pleasure to thank Emil J. Martinec, David R. Morrison, Sergej Moroz, and Nicholas Read for discussion.  This work is supported, in part, by DOE grant DE-FG02-13ER41958, NSF grant DMR-0820054 and a Simons Investigator grant from the Simons Foundation.



\appendix

\section{Superfluid Correlation Functions}
\label{App:SF}
Consider perturbations about a flat background with only an electromagnetic chemical potential turned on,
\eq{
a= \pi n_0 (x dy - y dx) + \epsilon a_\mu dx^\mu, g_{\mu\nu} = \eta_{\mu\nu} + \epsilon h_{\mu\nu},~ A= \mu_A  dt  + \epsilon A_\mu dx^\mu.
}
The two linearized equations of motion in the $a_t=0$ gauge are:
\begin{align}
\label{eq:SF_EOM}
0=& 2 q \varepsilon^{ij}F_{jt}+\frac{\kappa}{2}\left(\varepsilon^{ij}\varepsilon^{kl}\d_l(\d_j h_{kt}-\d_t h_{jk})
+\d_t\d^i h_{tt}-2\d_t^2 h_{it}
+\varepsilon^{ij}\d_t^3 a_j / n_0 \pi\right)\notag\\
&+\varepsilon'(n_0)\left(\varepsilon^{ij}\d_j h_{tt} - 2 \varepsilon^{ij}\d_t h_{jt}
 -\d_t^2 a^i / n_0 \pi \right)\notag\\
&+ \varepsilon''(n_0)n_0\left(\varepsilon^{ij}\delta^{kl}\d_j h_{kl}
+\varepsilon^{ij}\varepsilon^{kl} \d_l\d_j a_{k} / n_0 \pi \right)\notag\\
&+\frac{2}{n_0}\zeta(n_0) \left( \d_t\d^i h_{tt} - 2 \d_t^2 h_{it}
+ \varepsilon^{ij}\d_t^3 a_j /n_0 \pi \right)\notag\\
&+ \zeta'(n_0)\left(2 \varepsilon^{ij}\varepsilon^{kl}\d_l\d_j h_{kt}
+\delta^{jk}\d_t \d^i h_{jk} - \varepsilon^{ij}\d_t \delta^{kl}\d_k\d_l a_j /n_0 \pi \right),
\end{align}
where the indices are all taken to be spatial and $h=\eta^{\mu\nu}h_{\mu\nu}=-h_{tt}+h_{xx}+h_{yy}$.  Solving these equation at first order in momentum expansion and plugging back into the action, we can read off the various two point functions of the background fields $A_\mu$ and $h_{\mu\nu}$.

The electromagnetic response can be characterize by the polarization tensor $\Pi^{\mu\nu}$:
\begin{equation}
	j^\mu(\omega,\mathbf{k})=\Pi^{\mu\nu}(\omega,\mathbf{k}) A_\nu(\omega,\mathbf{k}).
\end{equation}
Following \cite{Son:2013}, we parameterize  $\Pi^{\mu\nu}$ using three functions $\Pi_{0,1,2}(\omega,\mathbf{k})$:
\begin{align}
\Pi^{00}=&k^2 \Pi_0,\notag\\
\Pi^{0i}=& \omega k_i \Pi_0- i \varepsilon^{ij}k_j \Pi_1, \hspace{15pt}
\Pi^{i0}= \omega k_i \Pi_0+ i \varepsilon^{ij}k_j \Pi_1,\\
\Pi^{ij}=&\omega^2 \delta^{ij}\Pi_0 + i \varepsilon^{ij}\omega \Pi_1 + (k^2\delta^{ij}-k^ik^j)\Pi_2.\notag
\end{align}
We calculate:
\begin{align}
	\Pi_0=&\frac{  q^2 }{\mu}\frac{n_0+\cO(k^2)}{c_s^2 \mathbf{k}^2-\omega^2},\\
	\Pi_1=&-\frac{q^2}{\mu^2}\frac{n_0 \zeta'(n_0) \mathbf k^2 - (2 \zeta(n_0)+\kappa n_0/2) \omega^2+\cO(k^4)}{c_s^2 \mathbf{k}^2-\omega^2},\\
	\Pi_2=&-\frac{q^2}{\mu}\frac{c_s^2 n_0+\cO(k^2)}{c_s^2 \mathbf{k}^2-\omega^2}.
\end{align}
As expected, we see that the the new terms added at order $O(p)$ to the action only affect $\Pi_1$ to the order we work to. 

We next look at electromagnetic response to metric perturbations. We find:
\begin{align}
j^t(\omega,\mathbf{k})=&q n_0 \delta(\omega,\mathbf k)+\frac{q}{c_s^2 \mathbf k^2 - \omega^2} \bigg[\frac{n_0}2\Big(\mathbf k^2(\tilde h_{tt}+c_s^2\delta^{kl} \tilde h_{kl})+2\omega k^i \tilde h_{it}\Big)\\
&\hspace{150pt}-\frac{i n_0 \kappa}{4 \mu}\varepsilon^{ij}\delta^{kl}k_ik_k\big(k_l \tilde h_{jt} 
+\omega \tilde h_{jl}\big)\bigg]\notag,\\
j^i(\omega,\mathbf{k})=&\frac{q}{c_s^2 \mathbf k^2 - \omega^2} \bigg[\frac{n_0}2  \Big(2  c_s^2 \varepsilon^{ij} k_j \varepsilon^{kl}  k_k \tilde h_{lt}+c_s^2 k^i  \omega \delta^{jk} \tilde h_{jk}
+k^i \omega \tilde h_{tt}+2\omega^2{\tilde h^i}_t\Big)\notag\\
&-\frac{i n_0 \kappa }{4\mu}\Big(c_s^2 \varepsilon^{ij}k_j\left(\varepsilon^{kl}\varepsilon^{mn}k_l k_m \tilde h_{kn}
+ \delta^{kl}k_k k_l \tilde h_{tt}+2 \omega k^k\tilde h_{kt}+\omega^2 \delta^{kl}\tilde h_{kl}\right)\notag\\
&\hspace{60pt}+\omega \varepsilon^{jk} k_j \left(k^i  \tilde h_{kt}+\omega {\tilde h^i}_k\right)\Big)\\
&-\frac{i}{\mu}\Big(c_s^2 \zeta(n_0)-\frac{n_0}{2}\zeta'(n_0)\Big)\varepsilon^{ij}k_j \Big(\delta^{kl}k_kk_l\tilde h_{tt}
+2\omega k^k \tilde h_{kt}+\omega^2 \delta^{kl}\tilde h_{kl}
\Big)\bigg]\notag,
\end{align}
where $\tilde h_{\mu\nu}=\tilde h_{\mu\nu}(\omega,\mathbf k)$.

And finally the gravitational response is:
\begin{align}
T^{tt}(\omega,\mathbf{k})=&\varepsilon(n_0)\big(\delta(\omega,\mathbf{k})+\tilde h_{tt}\big)
+\frac{1}{c_s^2 \mathbf k^2 - \omega^2} \bigg[\frac{n_0\mu}2\delta^{ij}\big(k_ik_j \tilde h_{tt}+\omega^2 \tilde h_{ij}+2\omega k_i \tilde h_{jt}\big)\notag\\
&-\frac{i n_0 \kappa}{4 } \varepsilon^{ij}\delta^{kl}k_ik_k\big(k_l\tilde h_{jt}+\omega \tilde h_{jl} \big)\bigg]\\
T^{ti}(\omega,\mathbf{k})=&- \tilde h^{ti}\varepsilon(n_0)+\frac{\frac12}{c_s^2 \mathbf k^2 - \omega^2} \bigg[
n_0 \mu k^i \Big(\omega h_{tt}+c_s^2 \delta^{jk}(2k_j\tilde h_{kt}+\omega \tilde h_{jk}) \Big)\\
&\hspace{-52pt}-\frac{i n_0 \kappa}{4}\Big(\varepsilon^{ij}k_j
(\delta^{kl}k_k  k_l \tilde h_{tt} -2 c_s^2 \varepsilon^{mn}\varepsilon^{kl} k_l k_m \tilde h_{kn})+2 \omega \varepsilon^{ij}\delta^{kl}k_kk_l \tilde h_{jt}
-\omega^2 \varepsilon^{jk}{\tilde h^i}_jk_k+\omega^2\varepsilon^{ij}k^k\tilde h_{jk}\Big)\bigg]\notag\\
T^{ij}(\omega,\mathbf{k})=&\big(n \mu-\varepsilon(n_0)\big)\big(\delta^{ij}\delta(\omega,\mathbf{k})-\tilde h^{ij}\big)
\notag \\
&+\frac{1}{c_s^2 \mathbf k^2 - \omega^2} \bigg[
	\frac{c_s^2 n_0 \mu}{2}
	\delta^{ij} \delta^{kl}\Big(k_kk_l\tilde h_{tt}+2\omega k_k \tilde h_{lt}
	+\omega^2\tilde h_{kl}\Big)\\
&-\frac{i n_0 \kappa}{4}\Big(\frac12
	(\varepsilon^{ik}\delta^{jl}+\varepsilon^{jk}\delta^{il})
	\big(\omega k_k k_l \tilde h_{tt}
	+2\omega^2 k_k \tilde h_{lt}+\omega^3 \tilde h_{kl}\big)\notag\\
&\hspace{180pt}+c_s^2 (\varepsilon^{ik}\varepsilon^{jl}+\varepsilon^{jk}\varepsilon^{il})
		\varepsilon^{mn}k_k k_m
		(k_l \tilde h_{nt}+\omega \tilde h_{ln})
	\Big)
\bigg],\notag
\end{align}
where again, $\tilde h = -\tilde h_{tt}+\tilde h_{xx}+\tilde h_{yy}$.

\addcontentsline{toc}{section}{Bibliography}
\bibliographystyle{JHEP}
\bibliography{WenZeerefs}

\end{document}